\documentclass[usenatbib]{mnras}

\usepackage{newtxtext,newtxmath}

\usepackage[T1]{fontenc}
\usepackage{ae,aecompl}

\usepackage{graphicx}	
\usepackage{amsmath}	
\usepackage{booktabs}
\usepackage{booktabs}
\usepackage[table,xcdraw]{xcolor}
\usepackage{soul}
\usepackage{float}
\usepackage{ulem}
\usepackage{color}           

\usepackage{multicol} 

\title[Cycle dependency of the quasi-biennial oscillation]{Cycle dependency of a quasi-biennial variability in the solar interior}

\author[T. Mehta et al.]{
T. Mehta,$^{1}$\thanks{E-mail: t.mehta.1@warwick.ac.uk}
K. Jain,$^{2}$
S. C. Tripathy,$^{2}$
R. Kiefer,$^{1,3}$
D. Kolotkov,$^{1}$
and A.-M. Broomhall$^{1,4}$
\\
$^{1}$ Centre for Fusion, Space and Astrophysics, Department of Physics, University of Warwick, Coventry, CV4 7AL, UK\\
$^{2}$ National Solar Observatory, 3665 Discovery Drive, Boulder, CO 80303, USA\\
$^{3}$ Leibniz-Institut f\"ur Sonnenphysik (KIS), Sch\"oneckstra\ss e 6, 79104, Freiburg, Germany\\
$^{4}$ Centre for Exoplanets and Habitability, University of Warwick, Coventry CV4 7AL, UK}

\date{Accepted XXX. Received YYY; in original form ZZZ}

\pubyear{2022}

\begin{document}
\label{firstpage}
\pagerange{\pageref{firstpage}--\pageref{lastpage}}
\maketitle

\begin{abstract}
We investigated the solar cycle dependency on the presence and periodicity of the Quasi-Biennial Oscillation (QBO). Using helioseismic techniques, we used solar oscillation frequencies from the Global Oscillations Network Group (GONG), Michelson Doppler Imager (MDI) and Helioseismic $\&$ Magnetic Imager (HMI) in the intermediate-degree range to investigate the frequency shifts over Cycles~23 and 24. We also examined two solar activity proxies, the $F_{10.7}$ index and the \ion{Mg}{ii} index, for the last four solar cycles to study the associated QBO. The analyses were performed using Empirical Mode Decomposition (EMD) and the Fast Fourier Transform (FFT). We found that the EMD analysis method is susceptible to detecting statistically significant Intrinsic Mode Functions (IMFs) with periodicities that are overtones of the length of the dataset under examination. Statistically significant periodicites, which were not due to overtones, were detected in the QBO range. We see a reduced presence of the QBO in Cycle~24 compared to Cycle~23. The presence of the QBO was not sensitive to the depth to which the p-mode travelled, nor the average frequency of the p-mode. The analysis further suggested that the magnetic field responsible for producing the QBO in frequency shifts of p-modes is anchored above approximately 0.95~R$_{\sun}$.

\end{abstract}



\section{Introduction}

Solar activity proxies such as the number of sunspots, solar radiation levels, and other quantities associated with the solar magnetic field fluctuate over an approximately 11\,yr period in what we call the Solar or Schwabe cycle. In addition to this largely periodic cycle, there is evidence of an oscillation of a shorter time scale known as the quasi-biennial oscillation (QBO), which is responsible for the double peak behaviour seen in some activity proxies at solar maximum. 
The periodicity of the QBO varies between $\leq$1 to $\sim$6\,yrs and is poorly defined while the amplitude of the QBO is known to be modulated by the solar cycle.  

QBO observations have been reported in a number of indices of solar origin such as the $F_{10.7}$ index, but have also been seen in data from the interplanetary magnetic field, solar energetic particles and other sources \citep[see][and references therein for a full review on observational aspects of the QBO]{2014Bazilevskaya}. Evidence of QBO-like signals have been seen in solar rotational rate residuals at near-surface depths \citep{2021InceogluB}. \citet{2017Broomhallb} showed that the amplitude of the QBO is highly correlated with sunspot number (SSN) and that the relative amplitude of the QBO to the 11\,yr cycle is constant across various activity proxies. The process generating the QBO is still not well understood. Several mechanisms have been proposed, and it is hoped through the use of modelling \citep[see][and references therein for a summary of suggested generation mechanisms]{2019Inceoglu} and observational data we may be able to better constrain the cause of the QBO. 

Following observations by \citet{1962Leighton} and \citet[][]{1962Evans} the field of helioseismology was born, centred on the study of resonant waves in the solar interior. Convective motions beneath the solar surface give rise to acoustic waves for which a pressure differential is the primary restoring force. Therefore these waves are known as `pressure' or `p-modes'. These modes are stochastically excited and inherently damped and may be observed through Doppler velocity or intensity images taken at the photosphere, among other methods. As the modes propagate through the solar interior, p-modes have been used in many studies to infer the conditions in the solar interior.

The periodicities of these p-modes, at approximately 5~minutes, vary slightly over the solar cycle, reaching their maximum frequencies in tandem with the solar activity maximum \citep[][]{1985Woodard}. This allows us to use frequency shifts of p-modes as tools to probe solar activity \citep[][]{2002Dalsgaard}. Solar oscillation frequencies have previously been used to examine the properties of the quasi-biennial oscillation as seen in \citet{2012Simoniello, 2013Simoniello,2009Broomhall, 2012Broomhall, 2010Fletcher}. Both \citet[][]{2013Tripathy} and \citet[][]{2013Simoniello} examined intermediate degree modes using GONG data over Cycle~23 using wavelet analysis and found evidence of oscillations in frequency shifts with a period of approximately 2\,yrs. \citet[][]{2012Broomhall} examined the QBO using low degree modes from Sun-as-a-star observations and showed a weak dependence of the QBO amplitude on mode frequency. 

A better understanding of the QBO has wide-reaching impact. For example, the solar wind is in part accelerated and driven by the solar magnetic field and affects us through its interactions with satellites, our energy grid, etc. Due to our incomplete understanding of the generation of the solar magnetic field, we are unable to fully predict its behaviour. By determining where and how the QBO is being generated we can refine our models of the solar magnetic field which will enable us to gain a fuller picture of the magnetic behaviour of the Sun and work towards mitigating some of the solar wind associated risks. The impacts of QBO research extend beyond our own solar system as there have also been observations of oscillations with mid-cycle periodicities on other fast-rotating stars \citep[see Figure 1 in][]{2007Bohm}, and in some star systems multiple periodicities are thought to have been seen. It is likely the phenomena responsible for these mid-cycle oscillations is similar to that governing the QBO and so the study of our Sun's QBO should lead to a better understanding of other solar-like star systems. 

The QBO is by definition quasi-periodic and therefore we expect to see a period drift in its profile. This makes it poorly suited to Fourier-based analysis techniques which anticipate a stationary sinusoidal input. Techniques which do not restrict the input to be strictly sinusoidal, e.g. wavelet analysis, have been previously applied to investigations involving the QBO  \citep[see][]{2012Simoniello}. However, the accuracy of wavelet analysis struggles in noisy conditions and so it may not be well suited for identifying quasi-periodic oscillations in the presence of both white and coloured noise which we see in solar data. In this study we make use of Empirical Mode Decomposition (EMD) on intermediate degree p-modes to obtain statistically significant periodicities in the QBO. EMD has previously been applied in QBO studies e.g. on cosmic ray intensity data wherein oscillations with periods in the QBO range were identified \citep[][]{2012Vecchio}, and on rotation rate residuals in Cycles~23 and~24 to investigate the presence of QBO-like signals \citep[][]{2021InceogluB}. Ensemble Empirical Mode Decomposition (EEMD), a technique similar to EMD, has also been used on data from Birmingham Solar Oscillations Network \citep[BiSON;][]{1996Chaplin, 2016Hale} where the study was confined to low degree modes \citep{2015Kolotkov}. 
 
In this paper we analyse p-modes using spatially resolved data obtained from the Global Oscillations Network Group {\citep[GONG;][]{1996Harvey, 2021Jain}} and a combination of data from the Michelson Doppler Imager \citep[MDI;][]{1995Scherrer} onboard the Solar and Heliospheric Observatory (SoHO) and the Solar Dynamics Observatory's (SDO) Helioseismic and Magnetic Imager \citep[HMI;][]{2012Scherrer} covering solar Cycles~23 and 24. The observed solar wavefield can be decomposed by projecting it onto spherical harmonics. In a power spectrum of the resulting time series, acoustic p-mode frequencies of individual harmonic degree ($\ell$), radial order ($n$), and azimuthal order ($m$) can be identified. The use of spatially resolved data allows us to select our choice of modes to investigate for parameters of interest, such as the depth to which the modes propagate. In this study, we primarily focus on the intermediate degree modes, where $\ell$ varies between 3 $\le\ell\le$ 150. We further examine solar activity proxies to search for QBO-like behaviour, specifically in the $F_{10.7}$ index and Bremen Composite \ion{Mg}{II} index, both of which contain high-quality data for several solar cycles. 
\newline

The structure of this paper is as follows. Section~\ref{sec:data} describes the data selection and processing methods. Section~\ref{sec:analysis} describes the analysis methods used, largely detailing the technique of EMD. Results of the analysis of GONG and MDI/HMI data over Cycles~23--24 are presented in Section~\ref{sec:results} alongside a comparison of analysis techniques, and the analysis of solar activity proxies over Cycles~21--24 is detailed in Section~\ref{sec:results2122}. Finally, the conclusions are summarised in Section~\ref{sec:conclusion}.

\section{Setting up the data}
\label{sec:data}
\subsection{Oscillation frequencies}
\label{freq}
\subsubsection{GONG}
\label{GONGdata}

Since 1995, GONG has provided mode frequencies produced by spatially resolved Doppler-velocity measurements via the GONG standard pipeline {\citep{1996Hill}}. For this analysis, we use the $m$-averaged frequencies for a given $\ell$ and $n$ mode, labelled as \textit{mrv1y}\footnote{\url{https://gong2.nso.edu}}. The \textit{mrv1y} files consist of 108-day overlapping datasets, centred on a 36-day GONG month. The $m$-averaged frequencies $\nu_{n,\ell}$ are determined by 
\begin{equation}
    \nu_{n,\ell, m} = \nu_{n,\ell} + \sum_{i} c_{i,n,\ell}   \gamma_{i,\ell}(m),
\end{equation}
where $\gamma_{i,\ell}(m)$ are orthogonal polynomials defined in Eqn. 33 of \citet{1991Ritzwoller} and $c_{i,n,\ell}$ are the Clebsch–Gordon splitting coefficients.

We examine the frequencies for all $n, \ell$ and create a time series $\nu_{n, \ell}(t)$ with 36 day cadence of the mode frequencies over both Cycles~23 and 24, and reject modes which are not present in all the datasets. Thus we analyse only the remaining modes known as `common modes' i.e. modes that have a well defined frequency for every GONG month in the time period investigated. We then calculate the frequency \textit{shift} time series for each common mode as $\delta\nu_{n,\ell}(t)$ where $\delta\nu_{n,\ell}(t) = \nu_{n,\ell}(t) - \bar{\nu}_{n,\ell}$ and $\bar{\nu}_{n,\ell}$ is the average mode frequency. The average mode frequency is the weighted mean frequency of a mode across the entire time span under consideration, where the weights are given by the inverse of the uncertainty on the fitted mode frequency. The value of $\bar{\nu}_{n,\ell}$ will, therefore, be slightly different depending on whether we are considering the Cycle 23, Cycle 24 or both. 

Every mode has an associated mode inertia which has a dependence on the modes frequency $\nu_{n,\ell}$ and harmonic degree $\ell$. We correct for this dependence using a mode inertial scaling factor $Q_{n,\ell}\epsilon_{n,\ell}$,  where $Q_{n,\ell}$ is the inertia ratio defined in \citet{cox1991solar} which suppresses the frequencies' harmonic degree dependence. The $\epsilon_{n,\ell}$ term acts to reduce the impact of the mode frequency. For discussions on the scaling factors and for examples of scaled frequency shifts where the frequency and degree dependence has been predominantly removed, the reader is referred to Figure~3 in \citet{2017Broomhalla} or Figure~1 in \citet{2001Chaplin}.

We obtain the corrected frequency shifts and trim the GONG datasets according to the procedure laid out in Section~\ref{sec:Trim} which resulted in 474 (with $1\le n \le 19$, 19 $\le \ell \le$ 147) and 743 (with 1 $\le n \le$ 20, 19 $\le \ell \le$ 145) common modes in the trimmed durations of Cycles~23 (GONG months 008--118 inclusive) and Cycle~24 (GONG months 139--244 inclusive). The start and end dates for corresponding datasets are given in Table \ref{tab:dates}. 

We sort the frequency shifts of the modes by mode frequency, choosing bins to have widths of  400$\,\mu$Hz overlapping by 200\,$\mu$Hz. We perform additional data sorting by calculating the `lower turning points', r$_{\text{ltp}}$, of the modes (where the lower turning point, which has a dependence on $\ell$ and $\nu_{n,\ell}$, is the depth at which the mode undergoes total internal refraction due to increasing pressure), following the standard method laid out in Model~S \citep[see][]{1996Dalsgaard}. The lower turning point bins have width of 0.05\,R$_{\sun}$, staggered by 0.01\,R$_{\sun}$ from 0.05\,R$_{\sun}$ < r$_{\text{ltp}}$ $\le$ 0.95\,R$_{\sun}$. We choose a depth resolution of the scale of the approximate width of the tachocline \citep{2009Howe} because the tachocline plays a key role in some QBO generation mechanisms \citep[for example, see][]{2010Zaqarashvili}. We sort all frequency shift time series $\delta\nu_{n,\ell}(t)$ into their corresponding bins, rejecting the bin if it contains fewer than three modes to reduce the impact of outliers. For a given bin, we calculate the weighted mean frequency, the average frequency shift time series and corresponding errors. The errors on the individual frequencies were produced by the standard GONG pipeline and used to weight the average frequency shift. The sorting of modes into bins and then averaging across the modes in a given bin to produce a single frequency shift time series results in 251 averaged modes to be assessed by EMD for Cycle~23, and 346 averaged modes for Cycle~24. 

\begin{table}
\caption{Trimmed durations for Cycles~21--24 for helioseismic and solar proxy datasets using DD/MM/YY notation. The inclusive time ranges for the below data were determined using the trimming method given in Section~\ref{sec:Trim} and are therefore shorter than the full durations of Cycles~23 and 24. All of the dates given contain the full duration of the datasets and are inclusive. The regions shaded in grey indicate where there is no data available. Note that the data availability for the \ion{Mg}{ii} index begins after Cycle 21 has started and MDI/HMI data pre-1999 is excluded due to the near loss of the SoHO spacecraft in 1998 which caused a gap in data collection.}
\label{tab:dates}
\begin{tabular}{ccccc}
\cline{1-4}
                      &                                                              & \multicolumn{2}{c}{Cycle}                           &  \\ \cline{3-4}
Dataset               & \begin{tabular}[c]{@{}c@{}}Cadence\\ {[}days{]}\end{tabular} & \textbf{21}              & \textbf{22}              &  \\ \cline{1-4}
\textbf{GONG}         & 36                                                           & \cellcolor[HTML]{C0C0C0} & \cellcolor[HTML]{C0C0C0} &  \\
\textbf{MDI/HMI}      & 72                                                           & \cellcolor[HTML]{C0C0C0} & \cellcolor[HTML]{C0C0C0} &  \\
\\
\textbf{10.7 cm flux} & 36                                                           & 05/01/77 -- 28/06/85      & 08/09/85 -- 09/12/95      &  \\
\textbf{10.7 cm flux} & 72                                                           & 11/01/76 -- 28/06/85      & 30/01/86 -- 28/09/95                        &  \\
\textbf{Mg II Index}  & 36                                                           & 20/11/78 -- 13/05/87       & 24/07/87 -- 12/05/94       &  \\
\textbf{Mg II Index}  & 72                                                           & 20/11/78 -- 20/12/86      & 04/10/87 -- 14/01/96      &  \\ 
\cline{1-4}
                      &                                                              & \multicolumn{2}{c}{Cycle}                           &  \\ \cline{3-4}
                      &                                                              & \textbf{23}              & \textbf{24}              &  \\ \cline{1-4}
\textbf{GONG}         & 36                                                           & 14/01/96 -- 22/12/06      & 12/12/08 -- 24/05/19      &  \\
Common modes &                                                           & 474      & 743      &  \\
\textbf{MDI/HMI}      & 72                                                           & 03/02/99 -- 11/12/08      & 07/12/09 -- 04/08/19      &  \\
Common modes &                                                           & 1304      & 1377      &  \\
\textbf{10.7 cm flux} & 36                                                           & 14/01/96 -- 22/12/06        & 12/12/08 -- 24/05/19      &  \\
\textbf{10.7 cm flux} & 72                                                           & 03/12/96 -- 16/05/07          & 07/10/07 -- 01/01/19                        &  \\
\textbf{Mg II Index}  & 36                                                           & 14/01/96 -- 07/10/07      & 28/02/08 -- 01/02/20       &  \\
\textbf{Mg II Index}  & 72                                                            & 14/01/96 -- 12/11/07      & 04/04/08 -- 21/11/19      &  \\ 
\cline{1-4}
\end{tabular}
\end{table}

\subsubsection{MDI/HMI}
\label{MDIdata}

The MDI instrument aboard SoHO was active from 1$\textsuperscript{st}$~May~1996 to 12$\textsuperscript{th}$~April~2011 and provided spatially resolved Doppler-velocity images with a cadence of 60 seconds where the mode frequencies were computed as 72-day cadence time series \citep{1995Scherrer,2015Larson, 2018Larson}. The successor to MDI is the HMI instrument aboard SDO which provides Dopplergrams with a cadence of 45 seconds, overlapped by five MDI 72-day datasets from its beginning of operations on 30$\textsuperscript{th}$~April~2010. These data can be found at the Joint Science Operations Center (JSOC) website\footnote{\url{http://jsoc.stanford.edu/ajax/lookdata.html}}. For convenience, the HMI dataset was chosen over the MDI data for the duration over which the data overlapped. Differences between the MDI and HMI mode frequencies are within 1-sigma (where $\sigma$ is the uncertainty in frequency determination) for the five overlapping 72-day datasets. The combined dataset uses the MDI data from 2$\textsuperscript{nd}$~February~1999 to 29$\textsuperscript{th}$~April~2010 and the HMI data from 30$\textsuperscript{th}$~April~2010 to 18$\textsuperscript{th}$~May~2020. This time range was later trimmed via the procedure discussed in Section~\ref{sec:Trim}. Following the method outlined for GONG data, we again sort these modes by both frequency and lower turning point, using the same frequency and lower turning point bins specified in Section~\ref{GONGdata}. The MDI/HMI combined data contained 1304 common modes (with 1\,$\le n \le$\,22, 5\,$\le \ell \le$\,150) in the trimmed durations of Cycle~23 and 1377 modes from MDI (with 1\,$\le n \le$\,22, 3\,$\le \ell \le$\,150) in Cycle~24. As with the GONG data, the weighted mean frequency shifts and corresponding errors were determined, using the uncertainties given by the MDI/HMI pipelines. Following sorting of these modes by lower turning points and average mode frequency, the frequency shift time series were averaged, where this averaging process decreases the total number of modes to be assessed for Cycles~23 and 24 to be 643 and 679 modes respectively. The datasets' start and end times are given in Table~\ref{tab:dates}.

\subsection{Solar activity proxies}
\subsubsection{\texorpdfstring{${F_{10.7}}$}{F10.7} index}
The $F_{10.7}$ index \citep{2013Tapping} measures the flux density of solar radio emission at a wavelength of 10.7\,cm and has been shown to correlate well with the frequency shifts of low \citep{2007Chaplin} and intermediate degree \citep{2009Jain} modes. It is a widely used proxy for solar activity as it correlates well with sunspot number, UV emissions and solar irradiance and daily observations are available since 1947. The data were obtained from the Canadian National Research Council website\footnote{\url{nrc.canada.ca/en}}, and were trimmed over the time periods given in Table~\ref{tab:dates}. The data were then averaged over 108-day bins, which each overlap by 36 days to mirror GONG's cadence, and 72-day independent bins to mirror MDI/HMI cadence and trimmed according to the same criteria laid out in Section~\ref{sec:Trim}. $F_{10.7}$ index emissions are measured in radio flux units (RFU), where 1\,RFU~=~10$\textsuperscript{-22}$\,Wm$\textsuperscript{-2}$Hz$\textsuperscript{-1}$.

\subsubsection{Bremen Composite \texorpdfstring{\ion{Mg}{ii}}{MgII} index}
First identified in \citet{1986Heath}, the Magnesium II
core-to-wing ratio of the \ion{Mg}{ii} Fraunhofer doublet at 280\,nm was found to be a good proxy for solar EUV irradiance. The Bremen composite (also known as the composite \ion{Mg}{ii} index)\footnote{\url{http://www.iup.uni-bremen.de/UVSAT/Datasets/MgII}} closely correlates with the Schwabe cycle \citep{2001Viereck} and has been constructed by combining the data collected by several instruments. The \ion{Mg}{ii} index has been once again trimmed over the time periods given in Table~\ref{tab:dates} via the trimming criteria discussed in Section~\ref{sec:Trim} and averaged over overlapping 108-day bins with 36 day cadence to mirror GONG's data, and 72-day independent bins to mirror MDI/HMI data.

\section{Analysis Methods}
\label{sec:analysis}

\subsection{Empirical Mode Decomposition}
\label{subsec:EMD}

\begin{figure}
\centering
\includegraphics[width=1.0\linewidth,trim={0cm 0.75cm 0cm 1.5cm}, clip]{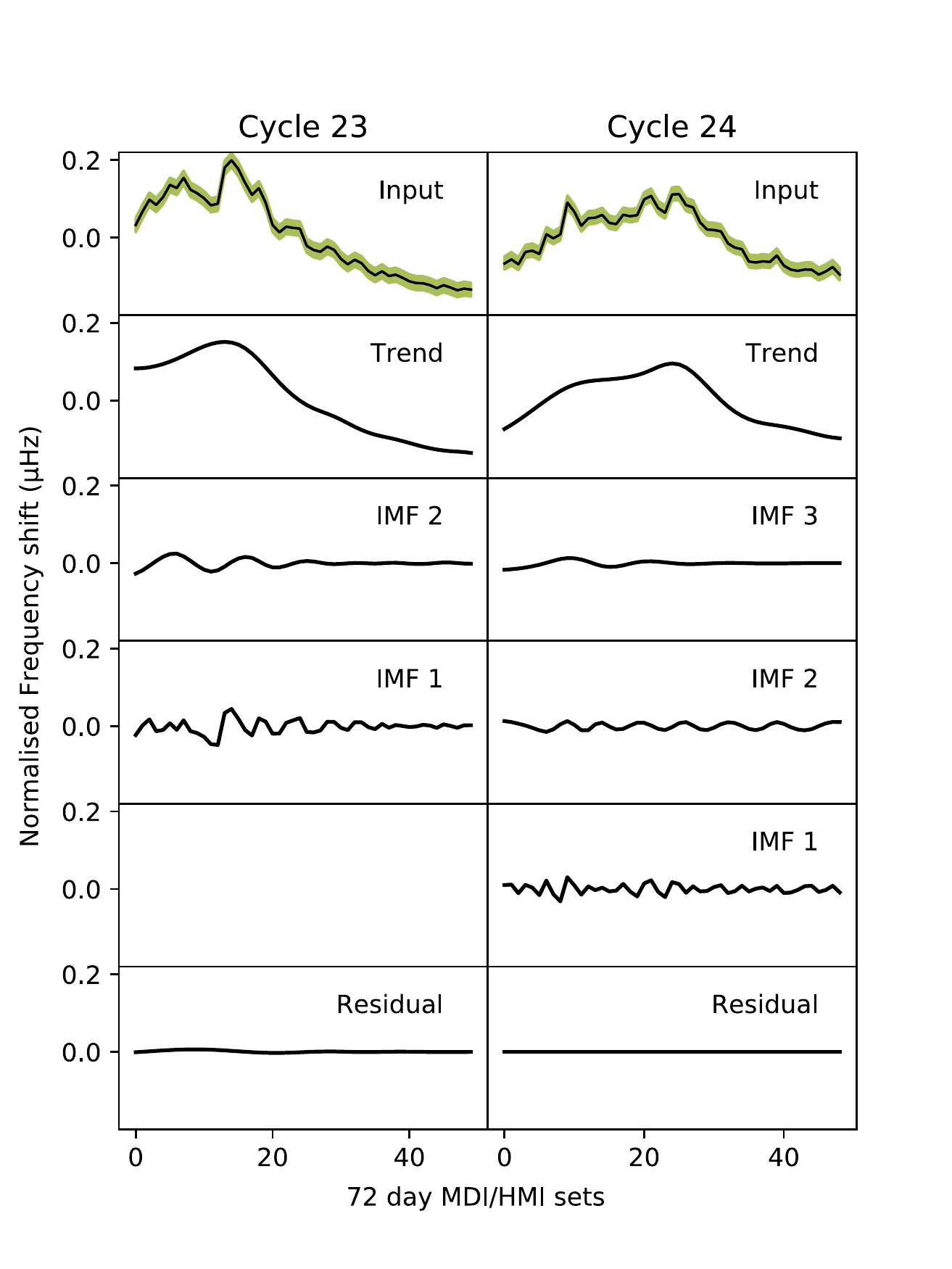}
\caption{ Averaged frequency shifts from modes with lower turning points between 0.7--0.75\,R$_{\sun}$ and frequency between 3200--3600\,$\mu$Hz during Cycle~23 (Top left) and Cycle~24 (Top Right) as measured by MDI/HMI. Frequency shifts have been normalised to have zero mean, with errors seen in green. Lower panels show the IMFs obtained from the decomposition from Cycles~23 (left) and 24 (Right). IMF~2 (with an average period of 721 days) from Cycle~23 and IMFs~2 (437 days) and 3 (889 days) from Cycle~24 are statistically significant above a 95\,\% confidence level. The IMFs with the shortest average periods from each cycle, (IMFs~1) are excluded from the analysis as they cannot be assessed with the confidence levels used.}
\label{fig:imfs}
\end{figure}

Empirical Mode Decomposition decomposes a dataset (such as a time series) into a set of unique Intrinsic Mode Functions (IMFs) (see Figure~\ref{fig:imfs} for an example of IMFs) which form a complete basis for the input signal \citep[see Section 5 of][and equations therein for a full derivation of the decomposition method]{1998Huang}. This is analogous to how Fourier analysis decomposes a signal into a series of sinusoids. However, IMFs are allowed to vary in period, amplitude and shape, which make them better suited to describe quasi-oscillatory processes which exhibit period drift and amplitude modulation. As the quasi-biennial oscillation is by definition quasi-periodic, it was imperative to make use of a technique that allows this variability in period. In EMD, each IMF is extracted through a sifting process \citep[detailed in Section~4.5 in][]{2019Broomhall}. Figure~\ref{fig:imfs} shows the decomposition of two input signals into their respective trends, IMFs, and residues. The input signals (top panels) are the frequency shift time series of averaged modes with lower turning points between 0.7--0.75\,R$_{\sun}$ and frequency between 3200--3600\,$\mu$Hz during their respective cycles as measured by MDI/HMI. The panels below show the frequency shifts' associated trends (determined as the combination of the IMFs with periods greater than half the duration of the signal). The remaining panels show the other IMFs found during the sifting process, and finally the residual of the signals wherein no more IMFs can be extracted as no oscillatory behaviour can be observed. The signal from Cycle~23 (left) decomposes into only two IMFs, whereas the signal from Cycle~24 (right) produces three. 

Once the IMFs are obtained they are judged against a confidence level. The confidence level used here relies on the assumption that the energy of an IMF follows a $\chi^{2}$ distribution (with degrees of freedom as a free parameter to be obtained), which is true for all IMFs except for the IMF with the shortest weighted-average period. Therefore we exclude the IMF with the shortest period in the analysis. The trend of the data is automatically selected as the sum of the IMFs with average periods greater than half the duration of the signal.

\begin{figure}
\centering
\includegraphics[width=0.95\linewidth]{2_EMD_FOURIER_tt.pdf}
\caption{ (Top) EMD spectrum of frequency shift obtained from MDI/HMI data during solar Cycle~24 seen in the right panel of  Figure~\ref{fig:imfs}. IMFs are visualised as bullet points indicating the IMFs weighted average period in days against their spectral energy. The 95$\%$ confidence level is shown in red. The IMF with the lowest period is shown in black and is excluded from the analysis as it cannot be assessed with the confidence level given. The two statistically significant IMFs and their associated errors in period are shown in orange, with periods of 437 and 889 days. The trend of the data is obtained as the sum of IMFs with periods larger than half of the duration of the signal, here shown in blue. (Bottom) Fourier spectrum of the same frequency shift time series. The red solid line shown on the spectra indicates a 95$\%$ confidence level.\label{fig:imfs_emdspec}}
\end{figure}

Confidence levels are determined via the assumption that the signal exists in the presence of white and coloured noise. We may fit a frequency shift's Fourier power spectrum with a broken power law to determine the non-zero value of $\alpha$ which characterises the colour of noise present in the signal. This is because the Fourier power spectral density $S$ and frequency $f$ are related as follows
\begin{equation}
    S\propto f^{-\alpha}
\end{equation}
where parameter $\alpha$ denotes the colour of the noise. As the data is of solar origin, we expect the presence of not only white noise ($\alpha=0$) but also some amount of coloured noise ($\alpha>0$).  
For an IMF extracted by EMD, its total energy $E_{m}$ is related to its average period $P_{m}$ by the following equation;
\begin{equation}
    E_{m}{P_{m}}^{1-\alpha} = \textrm{constant}
\end{equation}
where $\alpha$ is as previously defined.  By obtaining a value for $\alpha$ we may assess the statistical significance of the IMFs \citep[see][and references therein for a full description on determining confidence levels in EMD spectra]{2016Kolotkov}. 

We used Monte-Carlo simulation to determine the confidence levels in the EMD spectra. We first created two sets of 250 sample time series, one containing white noise and the other coloured noise,  such that the standard deviation of each sample was equal to the average error of the frequency shift data (green error bars in the top panel of Figure \ref{fig:imfs}). Note that the amplitude of these errors does not vary substantially with time for any single frequency shift time series. The simulated time series' were run through EMD to produce further test-IMFs, where the test-IMFs were used as the baseline to determine whether the corresponding IMFs obtained from real data were significant or not. To find the confidence level for a specific IMF, the energies of the test-IMFs were calculated and their distribution fitted by a $\chi^{2}$ distribution where degrees of freedom was a free parameter to be obtained by the fitting. Then using both the average energy and the degrees of freedom for these test-IMFs we found the value corresponding to a 95$\%$ confidence level given a $\chi^{2}$ distribution. By using the error on the input frequency shift data to determine the confidence levels in the EMD spectrum, we ensured our error analysis is appropriate, and consistent across all datasets.

The top panel of Figure~\ref{fig:imfs_emdspec} shows an EMD spectrum, which is a visualisation of this process, here showing the results from analysing the frequency shift shown in the top right panel of Figure\ref{fig:imfs}. The bottom panel of Figure~\ref{fig:imfs_emdspec} shows the Fourier power spectrum of the same frequency shift for comparison, with no peaks significant above the 95\% confidence level shown in red. The EMD spectrum plots the IMF's average periods against their energies. The IMF's weighted-average period is taken as the period corresponding to the peak value of the IMF's global wavelet spectrum. It is important to emphasise here that the IMFs are non-stationary by nature and usually exhibit a period drift. Therefore assigning a value for a stationary period is solely for convenience and gives an indication of the weighted-average period, where the weighting comes from the relative amplitude of the signal. 

The error on the IMF's period was found by taking the width between the period corresponding to maximal power and the periods which correspond to half maximum of the global wavelet spectrum (see bottom panel of Figure~\ref{fig:3panel}). The aforementioned non-stationary behaviour of many of the observed IMFs distorts the distribution of power in a global wavelet spectrum. This effect is compounded by the limited resolution of the wavelet which causes broader peaks in period space for longer periods. Therefore the resulting global wavelet spectra may naturally show a considerable spread in the period domain leading to large errors. We emphasise that the 95$\%$ confidence level used here and throughout the paper relates to the statistical significance of the IMF and the probability that it is not an artefact of noise. Therefore we may discuss an IMF with a given period as being above, e.g. a 95$\%$ confidence level, implying that there is a $5\%$ probability (or less) that the IMF is the result of noise in the signal. This does not tell us anything about the confidence of the weighted-period of the IMF. The routines used in this analysis can be accessed here.\footnote{\url{github.com/Sergey-Anfinogentov/EMD_conf}. We note that this project is still under development, and invite the reader to contact the author (d.kolotkov.1@warwick.ac.uk) directly for the most up to date versions of the technique and guidance.}

\begin{figure}
\begin{centering}
\includegraphics[width=0.95\linewidth]{3_GWS_tt.pdf}
\caption{
Normalised amplitude of a non-stationary IMF (Top) from MDI/HMI over the trimmed duration of Cycle~24 and its associated global wavelet spectrum (Bottom). The period associated with maximal power in the Global Wavelet Spectrum is indicated with a vertical orange line. The solid horizontal orange line shows the half-maximum value, used to determine the errors on the average period.} \label{fig:3panel}
\end{centering} 
\end{figure}

Returning to Figure \ref{fig:imfs}, IMF~2 from Cycle~23 exhibits QBO-like behaviour with an average period of 721 days, and further analysis showed it to be statistically significant above the 95$\%$ level. In Cycle~24, both IMFs~2 and 3 were again determined to be significant above the 95$\%$ level, with IMF~3 exhibiting a period of 889 days again falling in the range of expected QBO periodicities. IMF~2 from Cycle~24 had a period of 437 days, slightly above what we would expect for the annual oscillation.

\subsection{Fast Fourier Transform}
\label{sec:fft}

We also analysed a subset of datasets using a Fast Fourier Transform (FFT) to examine whether any QBO-like oscillations could be seen in a Fourier spectrum, and if so, whether they showed agreement with findings by EMD. The FFT was applied to detrended frequency shift time series that have already identified one or more statistically significant IMFs. The FFT assumes that the input signal is of stationary sinusoidal origin. As the QBO exhibits quasi-oscillatory behaviour by definition, an FFT of such a signal may cause spectral power to be spread over a number of bins in the frequency range to account for the spread in periodicity. This causes the spectral power to be distributed over a number of bins, reducing the chance of a given bin containing enough spectral power to be significant. Therefore, for signals with non-stationary and non-sinusoidal properties, such as the QBO, the FFT is a less effective tool in detecting periodicities. The statistical significance of peaks in the Fourier spectra were determined using the methodology described in \citet{2017Pugh}. Results of this analysis are discussed in Section~\ref{sec:results} for GONG and MDI/HMI data and in Section~\ref{sec:results2122} for analysis of solar activity proxies. The errors of significant periods found by the FFT were propagated using the functional approach \citep[See Chapter 4 in][for a description of the methodology used]{hughes2010measurements} and were taken to be the width of one frequency bin in frequency space. 

\subsection{Trimming the data}
\label{sec:Trim}

It is known that the amplitude of the QBO is modulated by the solar cycle and is greatest at solar maximum \citep{2014Bazilevskaya}. Therefore to increase the likelihood of observing the QBO we exclude solar minima by trimming the datasets. As the $F_{10.7}$ index is well correlated with GONG 108-day frequency shifts \citep[][]{2009Jain}, we use it to determine when solar activity rises above background levels.

We rebinned the $F_{10.7}$ index, averaging over overlapping 108-day bins with a 36-day cadence to match GONG months. These rebinned $F_{10.7}$ index data were then decomposed using EMD to produce a number of IMFs, and then the IMF with a period closest to 365~days was selected as the `annual oscillation'. This annual oscillation (IMF$_{365}$) was selected because its amplitude typically grows in magnitude towards solar maxima and decays towards solar minima and so it serves as a good tool for determining the bounds of solar maximum. In the case of the GONG data, IMF$_{365}$ has a period of 304$\genfrac{}{}{0pt}{2}{+79}{-67}$ days. We determined the dates where the amplitude of $(\textrm{IMF}_{365})^2$ is greater than 5$\%$ of this average maximum amplitude (the average of the three maximal amplitudes over the solar cycle) and trimmed the data to this duration.  This process excludes only the lowest amplitude values in the cycle, analogously trimming solar minima. This was repeated for solar Cycles~23 and 24, with $F_{10.7}$ rebinned to match the 72-day independent sets when used to trim the MDI/HMI data. The different rebinning of the $F_{10.7}$ index leads to a different IMF$_{365}$ profile obtained by EMD which in turn produces the different start and end dates for GONG and MDI/HMI data given in Table~\ref{tab:dates}.

Similarly we trimmed the solar activity proxies, covering data from Cycles~21--24. We performed trimming on each dataset twice; once using the overlapping 108-day bins with a 36-day cadence to match GONG months and again to match the 72-day independent sets used by MDI/HMI. This is to ensure the results from solar activity proxies can be directly compared to the results from helioseismic data. The trimming procedure was identical to the helioseismic trimming wherein the amplitude of IMF$_{365}$ obtained from the respective solar activity proxy was used to infer when the activity of the cycle is above background levels. We note that \ion{Mg}{ii} data only begins on 20$\textsuperscript{th}$ November 1978 and so a part of the rising phase of Cycle~21 is missing from the analysis. 

\section{Results}
\label{sec:results}

\subsection{Combined Cycles}
\label{subsec:results2324}

\begin{figure*}
\centering
\includegraphics[width=0.67\textwidth, trim={5cm 7cm 5cm 12cm}, clip]{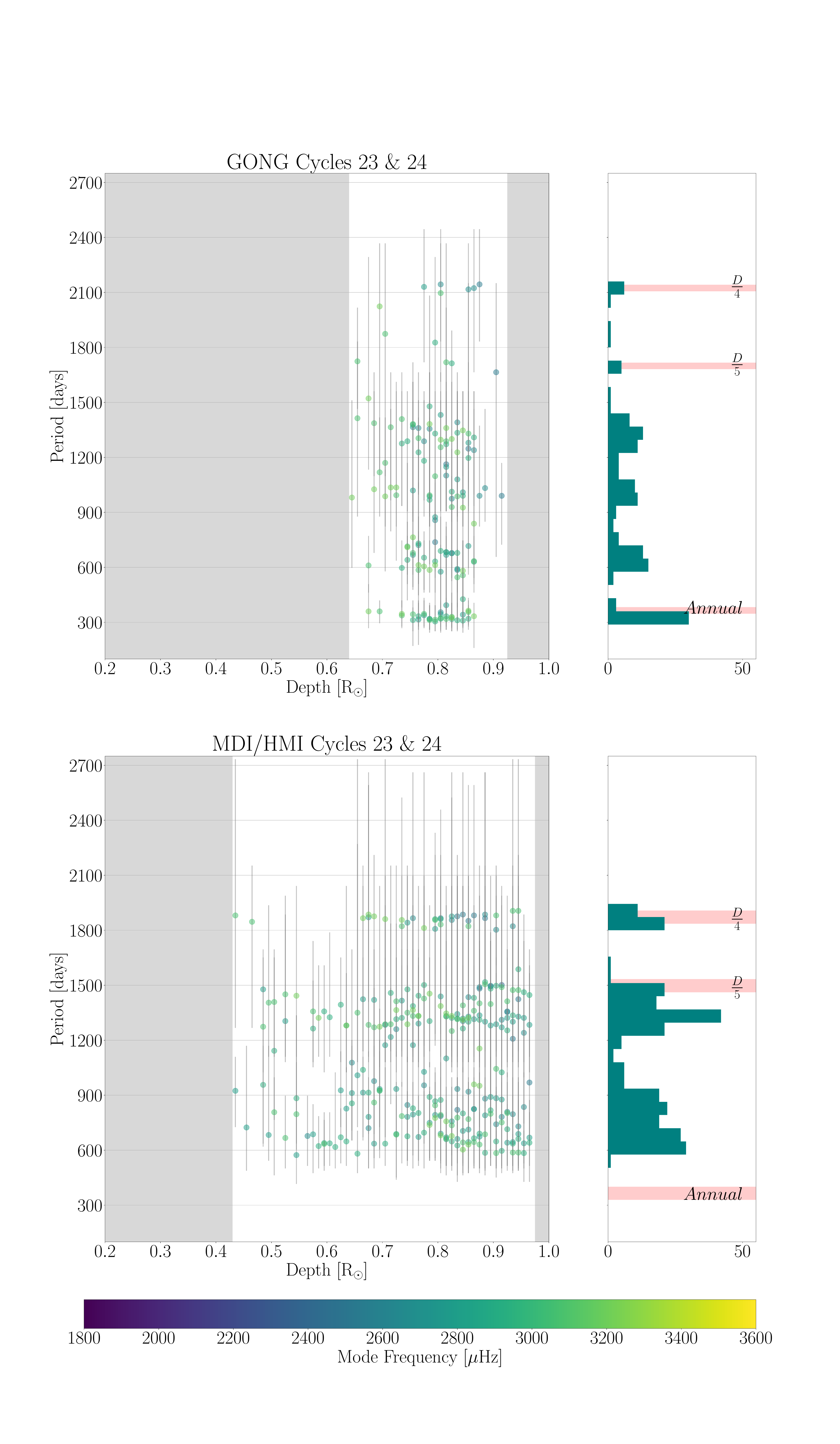}
\caption{Distribution of periods from statistically significant IMFs. The IMFs have been obtained from GONG data (top) and MDI/HMI data (bottom) from the combined duration of Cycles~23 and 24. The depth refers to the lower turning point in solar radius. The colour of the data points indicates the average p-mode frequency, and the error in period (found by the full width half maximum of the IMFs global wavelet spectrum) is indicated by grey error bars. The grey shaded region indicates depths over which  no data were available. The adjacent right panel shows the histogram of the distributions with bin-width equal to the cadence of the dataset. We include pink shaded regions (with width equal to twice the cadence of the dataset) as an eye-guide for the reader to visualise the approximate ranges over which overtones are expected, and the location of the annual oscillation.} The overtones are labelled as $\frac{D}{3}$,$\frac{D}{4}$, etc., where $D$ is the duration of the input signal. As discussed in Section~\ref{subsec:results2324}, six IMFs from MDI/HMI data with periodicities greater than 2400 days have been removed both from the scatter plot and its associated histogram (lower-left and lower-right panels) as their errors extend beyond the range plotted.  \label{fig:combined2324} 
\end{figure*}
Figure~\ref{fig:combined2324} shows the results of analysis for both GONG and MDI/HMI data over the time range spanning from the start of the trimmed duration of Cycle~23 to the end of the trimmed Cycle~24, as indicated in Table~\ref{tab:dates} for the respective datasets and contains the full duration of solar minima for Cycle 24. The figure shows the mode's lower turning point (in units of solar radius, R$_{\odot}$) against the period(s) of the mode's statistically significant IMF(s), given in units of days. We also indicate the mode's average frequency, in units of $\mu$Hz, as a colour bar. The grey shaded regions on the scatter plots (left) indicate where no high-quality data was available. The same data given in the scatter plots (left) are represented in an adjacent histogram (right), so that the reader may better visualise the distribution of the periods. The pink shaded regions, with a width of $\pm$ twice the cadence of the dataset, on the histograms show the ranges where overtones are expected as well as the region where we expect to see evidence of the annual oscillation. 

Overtones appear to be a result of the EMD technique, wherein statistically significant IMFs with periods of $\frac{D}{2}$, $\frac{D}{3}$, $\frac{D}{4}$ and infrequently $\frac{D}{5}$ (where $D$ is the duration of the input signal) are generated by the analysis. We shade in the approximate regions where $\frac{D}{3}$ to $\frac{D}{5}$ overtones are expected to aid the reader in differentiating overtones and other IMFs. A similar behaviour can be seen in Fourier analysis \citep[see e.g. Figures 4 and 5 in][]{2016Kolotkov}.

It is difficult to differentiate if a statistically significant oscillation with fewer than two cycles over its duration is part of the signal's trend and caused by noise in the data, or equally a physical phenomenon. Therefore we attribute IMFs with periods greater than $\frac{D}{2}$ to the general trend of the signal and suggest that these IMFs may be studied further over a longer duration and can therefore be analysed with higher accuracy.

The GONG results (top panel, Figure \ref{fig:combined2324}) show some evidence of banding, with clustering seen around 300--400, 500--800, 900--1200, 1200--1500 days, and a small number of points between 1700--2200 days where no dominant period can be discerned. The IMFs with periods in between 300--400 days can be attributed to the annual oscillation caused by the Earth's orbit, as GONG data is collected by ground-based instruments. The periods between 1700--2200 days are around the expected locations of the $\frac{D}{4}$ or $\frac{D}{5}$ overtones. We discuss the impact of overtones in more detail in Section \ref{subsec:resultsindep}. The other bands, clustered at 500--800, 900--1200, and 1200--1500 days may be attributed to the QBO. There is no clear dependence between periodicity and lower turning point or frequency, as the relative clustering of IMFs within the ranges of 0.7--0.9\,R$_{\sun}$ and 2600--3400\,$\mu$Hz, is largely due to selection bias of modes with 100$\%$ fill. 

The results from MDI/HMI (bottom panel of Figure \ref{fig:combined2324}) cover a larger depth range, with IMFs at depths upwards from 0.43\,R$_{\sun}$. The IMFs are clustered with reasonable scatter at 600--1000 days, 1200--1500, and 1900 days. We see no evidence of the annual oscillation in the MDI/HMI data. We consider IMFs with periods between 600--1000 to be exhibiting QBO-like behaviour. Once again there is no correlation between periodicity, lower turning point or mode frequency. The duration of the input signal, $D$, was 7487 days for this dataset so the resulting $\frac{D}{4}$ overtone is be expected at 1872 days - extremely close to the clustering of modes at 1900 days that is observed. The band between 1200--1500 days partly overlaps with the expected location of the $\frac{D}{5}$ overtone at 1497 days. However the range in periodicity of this band is considerably larger than the one at 1850 days, suggesting that either the overtones at $\frac{D}{5}$ exhibit more scatter, or that some of the IMFs in this range may be of real origin. For visual reasons we have excluded six datapoints from the combined MDI/HMI results (lower panels of Figure~\ref{fig:combined2324}) that correspond to IMFs with weighted average periods between 2472--2491 days. These average periods are within 25 days of 2496 days, which is the expected location of the $\frac{D}{3}$ overtone. We have omitted these data points in order to reduce the range on the vertical axis, for both the scatter plot (lower left) and its corresponding histogram (lower right) so that the reader may visualise the datapoints at lower periodicities with greater clarity.

\subsection{Independent Cycles}
\label{subsec:resultsindep}

\subsubsection{GONG}
\label{sec:analysissub}

\begin{figure*}
\begin{centering}
\includegraphics[width=0.7\textwidth, trim={5cm 7cm 5cm 13cm}, clip]{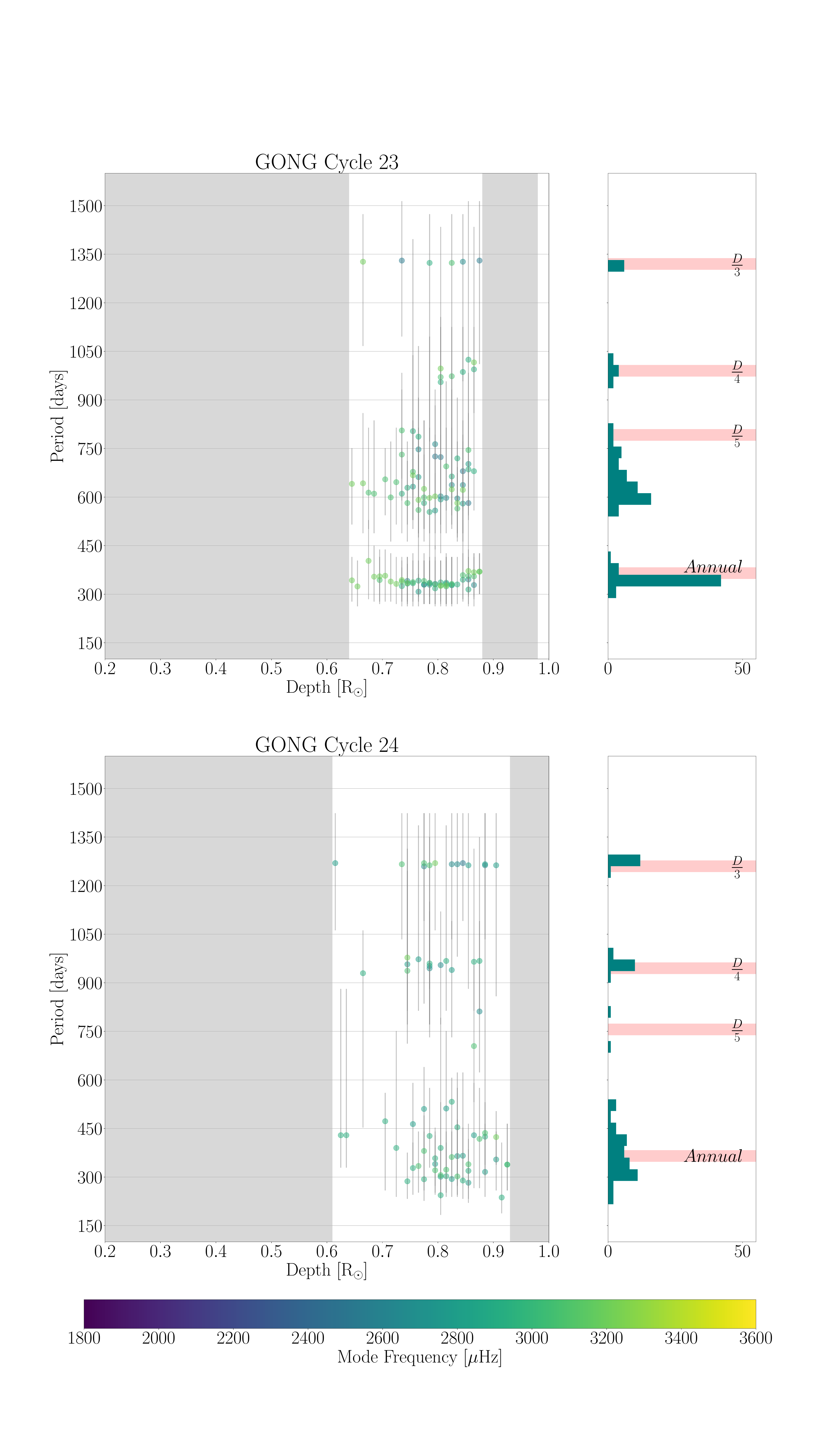}
\caption{\label{fig:GONG} Distribution of periods from statistically significant IMFs with IMFs obtained from Cycle~23 (top) and Cycle~24 (bottom) using GONG data.} Colours, shaded regions and axes have the same meaning as in Figure~\ref{fig:combined2324}. We do not expect all overtones to be present in all datasets and have included them as an eye-guide for the reader. Note that the vertical axis has different limits to that used in Figure~\ref{fig:combined2324}.
\end{centering}
\end{figure*}

The results of the analysis of modes obtained by GONG can be seen in Figure~\ref{fig:GONG} for the trimmed durations of Cycles~23 (top) and 24 (bottom). Following rebinning in frequency and lower turning point of the 474 common modes from Cycle~23, we obtain 251 averaged modes. Of these 91 produced at least one statistically significant IMF, equivalent to 36$\%$ of the sample. For Cycle~24, only 64 (18$\%$) of the 346 input averaged modes met this criterion.

For Cycle~23 the periods of the IMFs are roughly distributed across four bands: 300--400, 500--800, 950--1000, and 1300 days. The bands vary in appearance- e.g. the 500--800 day band is loosely defined and shows a lot of scatter in contrast to the bands at 950--1000 and 1300 days.

The 300--400 day band shows minimal scatter and is known to be an artefacts of the Earth's annual orbit around the Sun. The 950--1000 day and 1300 day bands are likely overtones of the input signal as they coincide with $\frac{D}{4}$ (990 days) and $\frac{D}{3}$ (1320 days). We also note that the bands are closely packed and show little scatter which are characteristic properties of an overtone. 

In contrast the 500--800 day band shows a lot of scatter and contains candidates for quasi-biennial oscillatory behaviour. The band shows some overlap with the location of the $\frac{D}{5}$ overtone which, if present, occurs at 790 days as the duration of the input signal is 3960 days. However the large range of periods covered by this band suggests some IMFs are from non-overtone origin.

There is no clear depth dependence beyond the lack of statistically significant IMFs with lower turning points smaller than 0.6\,R$_{\sun}$. We do not observe any correlation between the periods of IMFs and their helioseismic frequencies.

The data from the trimmed Cycle~24 seen in the bottom panel of Figure~\ref{fig:GONG} shows similar banding behaviour in the approximate ranges of 250--500, 950--1000 days, and 1275 days. The cluster of results in the 250--500-day range may contain some IMFs which are the results of the Earth's orbit. However, in contrast to the annual bands seen in the upper panels of Figures \ref{fig:combined2324} and \ref{fig:GONG}, the annual band seen in Cycle 24 covers a wide range of periods. It is possible that some IMFs with periods in the upper range of this cluster may be attributed to the QBO but there is no clear boundary between IMFs from the annual and QBO-like regime. The bands at 950--1000 and 1275 days show little scatter and coincide with $\frac{D}{3}$ (1260) and $\frac{D}{4}$ (945) where the duration of the signal $D$ is 3780 days. This suggests that these bands are overtones of the input signal. Due to the shorter duration of Cycle~24, the overtones are more densely packed than those in Cycle~23. A shorter duration naturally leads to more common modes, which in turn results in a greater population of overtones. However we see a weaker QBO presence in Cycle~24 than in Cycle~23 despite the increased number of common modes in Cycle~24. We observe 49 IMFs with periods in the range of 500--800 days which can be associated with the QBO for Cycle~23, with a roughly triangular distribution in its associated histogram, with a median value around 630 days. We contrast this to the same period range for Cycle~24, where we see no distinct peak with only 4 IMFs (a 92$\%$ decrease) with a weighted average period between 500--800 days, and a median value of 520 days. 

We further carried out FFT analysis on the independent helioseismic datasets. For both Cycle~23 and Cycle 24, the Schwabe cycle was infrequently detected above a confidence level of 95$\%$ by Fourier analysis (where a detection is deemed as a statistically significant peak with period 3650$\le P \le$ 4380 days). This was largely because only one or fewer Schwabe cycles were present in the signals due to the trimming. Although the Schwabe cycle dominated the signal in terms of amplitude, it did not exhibit perfect sinusoidal behaviour and therefore its power was distributed over a number of frequency bins. This is further discussed in Section~\ref{subsec:comparisonanalysis}. Therefore, we repeated this analysis with the detrended signal, with the trend being extracted by EMD, as the input signal for both Cycles~23 and 24 (and similarly for the results from MDI/HMI) in order to examine shorter periodicity oscillations, such as the QBO. We assessed 251 datasets for Cycle~23 and 346 for Cycle~24. 

Of the 251 input signals for the detrended datasets from trimmed Cycle~23, 102 (corresponding to 41$\%$) Fourier spectra had a peak between 333--363 days indicating the presence of the annual oscillation. We estimate the errors on these periods via the so-called functional approach. For the periods between 333--363 days, the corresponding errors to 1sf are -30, +40 days. There were 5 further detections outside this range, all at 666 days (with errors -100, +100 days) which are suggestive of QBO behaviour. Cycle~24 yielded fewer detections overall, in which only 32 of the 346 datasets (9$\%$) produced statistically significant results. The vast majority of these detections had periods between 214--385 days (errors -40, +40 days), with the only other detection at 428 days (error -40, +50 days). It is difficult to determine whether these are truly of solar origin or are indeed false positives due to the low number of detections. To improve confidence in this result we require either more data or an analysis tool which is better suited to detect quasi-periodic signals. These results suggest that solar oscillations with periods less than the Schwabe cycle were less likely to be seen in Cycle~24 compared to Cycle~23. 

\subsubsection{MDI/HMI}
\label{sub:mdires}

\begin{figure*}
\begin{centering}
\includegraphics[width=0.74\textwidth, trim={5cm 7cm 5cm 12cm}, clip]{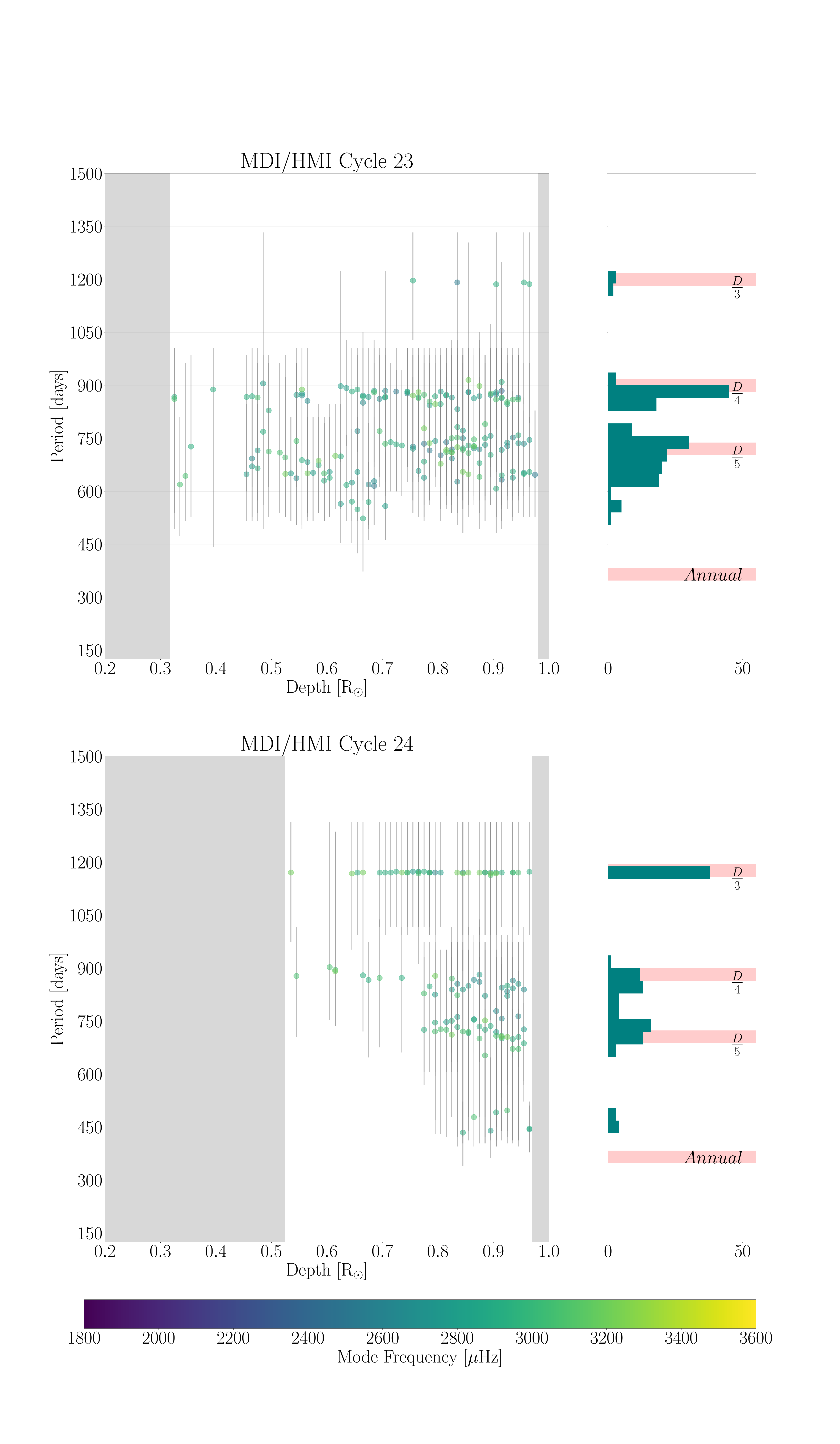}
\caption{\label{fig:MDIHMI} Distribution of periods from statistically significant IMFs with IMFs obtained from Cycle~23 (top) and Cycle~24 (bottom) using MDI/HMI data. Colours, symbols etc. are all as given in Figures~\ref{fig:combined2324} and \ref{fig:GONG}.}
\end{centering}
\end{figure*}

The MDI/HMI datasets have more than twice the number of common modes than the GONG datasets and so produced a greater number of IMFs over a greater depth range. The top panel in Figure~\ref{fig:MDIHMI} illustrates the distribution of periods from IMFs for the MDI/HMI modes detected over the trimmed duration of Cycle~23. There is evidence of banding at approximately 550--800, 800--900 and 1200 days. The $\frac{D}{5}$ overtone would be expected at approximately 705 days as the duration of the input signal for Cycle~23 was 3528 days. This lies within the 550--800 band. The 550--800 band also shows significant scatter which makes it difficult to differentiate overtones and potential QBO candidates. The $\frac{D}{4}$ and $\frac{D}{3}$ overtones are expected at 882 and 1176 days, which correlate with the bands at 800--900 and 1200 days. The periodicities of these QBO candidates again have no correlation with mode frequency. 

The results for the trimmed data of Cycle~24 are shown in the bottom panel of Figure~\ref{fig:MDIHMI}, producing bands at 400--500 days, 700--800 days, 800--900 days, and 1200 days. 

The IMFs with periods of 400--500 days have periodicities greater than what is usually attributed to the annual oscillation, consistent with the periodicities seen in the results from GONG data during Cycle~24. This may be evidence of QBO exhibiting a shorter period in Cycle~24 compared to Cycle~23. A number of IMFs with periods in the range of 550--800 days may also be associated with the QBO. For Cycle~23 we see 105 IMFs with weighted average periods between 550--800 days with a median of 700 days. In assessing the same range in Cycle~24 we see only 36 IMFs (a 66$\%$ decrease), with a median period of 730 days. This range does encompass the expected location of the $\frac{D}{5}$ (690-705 days) overtone for both cycles. The $\frac{D}{5}$ overtone (at 690 days, where the duration of the input signal is 3456 days) correlates reasonably well to the band at 700--800 days seen at depths above 0.75\,R$_{\sun}$ However, as the durations of both datasets were roughly the same we would expect the $\frac{D}{5}$ overtone to have similar prevalence in both datasets. Therefore the decrease in IMFs over this range is more likely to be of solar origin than a result of the analysis method. Again differentiating the origin of IMFs in this range poses a challenge as some of the IMFs in this band and the 800--900-day band may be considered QBO candidates. Both bands at 700--800 days and 800--900 days become more densely populated in the region corresponding to 0.75\,R$_{\sun}$ and above and exist across all frequencies for which low error data was available. The locations of the $\frac{D}{3}$ (1152 days) and $\frac{D}{4}$ (864 days) overtones correspond well to the bands at 1200 days and 800--900 days. Although the $\frac{D}{3}$ band is much more densely packed and better defined than its counterpart in the trimmed Cycle~23 MDI/HMI data, overtones corresponding to $\frac{D}{4}$ appear much more sparsely populated than its Cycle 23 counterpart.

Overall, the results from MDI/HMI show a similar pattern to those from GONG: there are proportionally fewer IMFs detected with periodicities in the QBO range in the trimmed Cycle~24 data than in the trimmed Cycle~23 data.

We examine the results from the FFT on the detrended signals both from the trimmed durations of Cycle~23 and 24, at a 95$\%$ confidence level. Fourier analysis of Cycle~23 yields 26 detections out of a possible 643 datasets (4$\%$) wherein oscillations at 367 days (with errors $\pm$ 40 days) were observed. A further 5 detections ($< 1\%$) of oscillations with periods of 183 days (with errors $\pm$ 10 days) were also seen. Interestingly, all of these detections occurred for lower turning points greater than 0.76\,R$_{\sun}$. This low number of overall detections from the FFT again may be in part due to the fact that the rising phase of the cycle was omitted so fewer annual cycles could clearly be seen. Out of 679 datasets in trimmed Cycle~24 data, there were 8 detections above the 95$\%$ confidence level, all at 1176 days (with errors -300, +600 days), aligning closely with the $\frac{D}{3}$ ($\sim$1200 day) band seen in the bottom panel of Figure~\ref{fig:MDIHMI}. This suggests that Fourier analysis also picked up overtones of the input signal. For Cycle~24 data in MDI/HMI, there were no detections of the annual oscillation. This follows on from the previous findings that Cycle~24 appears to show fewer detections of oscillations shorter than the Schwabe cycle. 

\subsection{Comparison of analysis techniques}
\label{subsec:comparisonanalysis}

As discussed in Sections~\ref{sec:analysissub} and \ref{sub:mdires}, the FFT procedure rarely detected the periodic annual oscillation in detrended frequency shift datasets. Therefore it is unsurprising that very few detections of the QBO were seen, as the QBO is expected to be not only lower in amplitude than the annual oscillation, but also its quasi-periodic behaviour spreads its power over several frequency bins. However, it is important to emphasise that FFT analysis was only performed on frequency shifts that had already identified at least one statistically significant IMF and did not investigate signals where no significant IMFs were detected. Therefore these results are useful for a comparison of the methods where EMD found statistically significant oscillations.

\begin{figure}
\includegraphics[width=\linewidth]{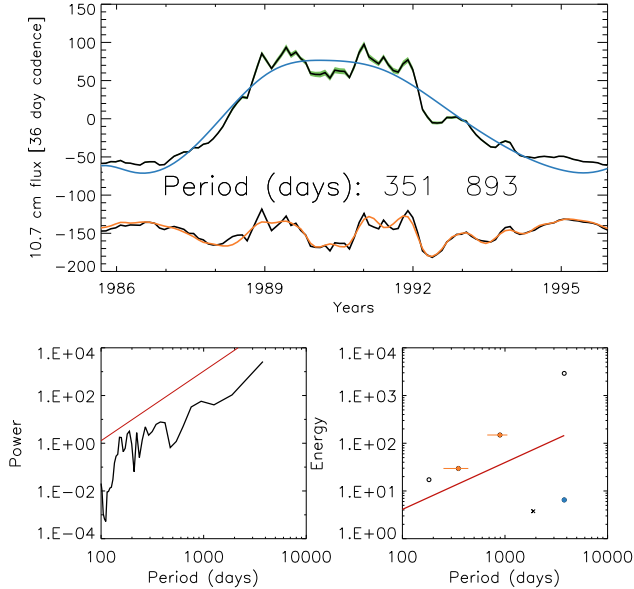}
\caption{ Analysis of F$_{10.7}$ data over Cycle~22 where data is averaged over 108-day bins, overlapping by 36 days. (Top) Upper curve shows the normalised F$_{10.7}$ data, with errors in green, superimposed with the IMF associated with the trend, shown in blue. Lower curve shows the detrended signal with the overlayed sum of two statistically significant IMFs. These IMFs have periods of 351$\genfrac{}{}{0pt}{2}{+88}{-100}$, 893$\genfrac{}{}{0pt}{2}{+163}{-221}$ days. (Lower left) FFT spectrum of the F$_{10.7}$ data, with  95$\%$ confidence level (solid red line). (Lower right) EMD spectrum of the F$_{10.7}$ data with IMFs visualised as bullet points, indicating their average periods and spectral energy. The 95$\%$ confidence level is shown by red line. The IMF with the lowest period cannot be assessed with the confidence level given and is excluded from the analysis. Statistically significant IMFs with energy over the confidence levels are seen in orange with the error in period shown as a horizontal orange line. The IMF associated with the trend is shown in blue and the residual is indicated by a cross. } \label{fig:radio22_36} 
\end{figure}

The distribution of power in period space for an IMF is explored in Figure~\ref{fig:3panel} where the top panel shows a statistically significant IMF from a MDI/HMI frequency shift time series in the trimmed duration of Cycle~24, with 2600\,<$ \nu_{n,\ell}\le$\,3000\,$\mu$Hz, 0.74\,< r$_{\text{ltp}}$ $\le$\,0.79\,$R_{\sun}$, with a weighted average period of 613$\genfrac{}{}{0pt}{2}{+360}{-134}$ days. Both amplitude modulation and a period drift can be seen in the mode profile. The lower panel shows the associated global wavelet spectrum of the mode which can be seen to have a significant spread across period space, from approximately 500--1000 days. A double-peak can be seen at roughly 600 and 900 days. This mode is symptomatic of many modes in this study and shows why it is inappropriate to be assigning a quasi-oscillatory signal a single value for period. By assigning this mode a single period, we erase the valuable time-dependant period drift behaviour (in this case we see that the period is greatest towards solar maximum of Cycle~24 in 2014 and then decreases during the decay phase), which may provide important information concerning the generation and behaviour of QBO signals. Figure~\ref{fig:imfs_emdspec} shows the EMD and Fourier spectra of a MDI/HMI frequency shift. Although we see evidence of statically significant oscillations in the EMD spectrum the Fourier spectrum shows no statistically significant peaks above the 95\,\% confidence level although we do see an enhancement of the Fourier power in the corresponding period ranges. We may attribute the lack of detection of significant peaks to spectral power leaking into adjacent bins due to the as the non-stationarity of the signal. This is why Fourier analysis is poorly suited to data with period drifts. The use of FFT in this analysis proved to be useful as an `eye-guide' only. The QBO also remained undetected in Fourier spectra of solar activity proxies (e.g. see Figure~\ref{fig:radio22_36}).

\section{Examining previous cycles using solar activity proxies}
\label{sec:results2122}

\begin{table}
\caption{The periodicities, given in days, of IMFs with statistical significance over a 95$\%$ confidence interval found by running EMD on data from solar activity proxies. IMFs that we do not attribute to overtones are emphasised in bold. Dashes indicate where no statistically significant IMFs were observed for the given dataset.}
\label{tab:proxyresult}
\begin{tabular}{ccccccl}
\cline{1-6}
                      &                                                              & \multicolumn{4}{c}{Cycle}                                                                         &  \\ \cline{3-6}
Dataset               & \begin{tabular}[c]{@{}c@{}}Cadence\\ {[}days{]}\end{tabular} & \textbf{21}              & \textbf{22}              & \textbf{23}          & \textbf{24}          &  \\ \cline{1-6} \\

\textbf{10.7 cm flux} & 36                                                           & \textbf{464$\genfrac{}{}{0pt}{2}{+96}{-151}$  }           & 
\textbf{351$\genfrac{}{}{0pt}{2}{+88}{-100}$ }                  & 
\textbf{308$\genfrac{}{}{0pt}{2}{+66}{-59}$   }            & 
961$\genfrac{}{}{0pt}{2}{+363}{-225}$                 &  \\

&  & &\textbf{893$\genfrac{}{}{0pt}{2}{+163}{-221}$}  &\textbf{ 669$\genfrac{}{}{0pt}{2}{+173}{-179}$} & & \\

& & & & 1332$\genfrac{}{}{0pt}{2}{+152}{-444}$      & & \\
\\
\textbf{10.7 cm flux} & 72                                                            & --                    & \textbf{286$\genfrac{}{}{0pt}{2}{+167}{-149}$  }                      & 747$\genfrac{}{}{0pt}{2}{+205}{-167}$                 & 
--                   &  \\
\\
\textbf{Mg II Index}  & 36                                                           & --                        & 839$\genfrac{}{}{0pt}{2}{+86}{-110}$                 &\textbf{ 584$\genfrac{}{}{0pt}{2}{+140}{-166}$   }                 & \textbf{448$\genfrac{}{}{0pt}{2}{+113}{-71}$   }               &  \\
& & & & &1246$\genfrac{}{}{0pt}{2}{+233}{-309}$    & \\
\\
\textbf{Mg II Index}  & 72                                                           & --                        & --                        & \textbf{648$\genfrac{}{}{0pt}{2}{+138}{-116}$}                   & 1081$\genfrac{}{}{0pt}{2}{+150}{-166}$                    &  \\ 

& & & & &         & \\
\cline{1-6}
\multicolumn{1}{l}{}  & \multicolumn{1}{l}{}                                         & \multicolumn{1}{l}{}     & \multicolumn{1}{l}{}     & \multicolumn{1}{l}{} & \multicolumn{1}{l}{} & 
\end{tabular}
\end{table}

In order to determine whether the QBO is a cycle-dependent process, we analyse its behaviour over multiple solar cycles using solar activity proxies since MDI/HMI and GONG have only been functional from the rising phase of Cycle~23. We turn to solar activity proxies which have high-quality data dating back several solar cycles. We use data from Cycle~21 through Cycle~24 in the \ion{Mg}{ii} index and $F_{10.7}$ index, and analyse it using EMD in a similar fashion to the oscillation frequency data. These datasets were also trimmed according to the method discussed in Section~\ref{sec:Trim} in order to exclude the time period correlating with low solar activity, where the QBO is less likely to be observed. The start and end dates of the solar activity proxy data are given in Table~\ref{tab:dates}. All data were normalised to have a zero mean. We again use a 95$\%$ confidence interval, and the periodicities of any statistically significant IMFs found through this method are stated in Table~\ref{tab:proxyresult}.
We account for errors on the solar proxies by taking the standard error on the mean $\sigma_{\bar{x}}$ where $\sigma_{\bar{x}} = \frac{\sigma}{\sqrt{n}}$. Here $\sigma$ is the standard deviation of the signal, and $n$ is the number of points we average over. We use this standard error to construct symmetrical error bars, shown in green colour in Figure~\ref{fig:radio22_36}.

Across both datasets we see only one detection of a statistically significant IMF for Cycle~21, found in the 36-day cadence $F_{10.7}$ data. This IMF had a weighted average period of 464 days, which is within the range we may attribute to the annual oscillation period. This period is similar to another detected in Cycle~24 which is discussed further below. 

Cycle~22 yields detections for three out of the four datasets, with only the 72-day cadence \ion{Mg}{ii} index data showing no statistically significant IMFs. We see IMFs with an average periods of 286 and 351 days (both from $F_{10.7}$ index data) which we may attribute to the annual oscillation. The remaining detections (seen in 36-day cadence for both $F_{10.7}$ and \ion{Mg}{ii} index data) have periods of 839 and 893 days, and are in the range commonly associated with the QBO. We examine if any of these detections lie close to the expected regions associated with overtones. The durations of 36-day cadence $F_{10.7}$ and 36-day cadence \ion{Mg}{ii} index data were 3744 and 2484 days, respectively. The detection from 36-day cadence \ion{Mg}{ii} index data at 839 days lies close to the $\frac{D}{3}$ overtone, at 828 days. However, the detection in 36-day cadence $F_{10.7}$ data at 893 days cannot be easily attributed to an overtone. Further it correlates well with the other detection in this range suggesting that the IMFs with periods in this range may be of real solar origin.

Cycle~23 shows evidence of the annual oscillation for the 36-day rebinned $F_{10.7}$ data, yielding a detection with an average period of 308 days. We also see detections of IMFs with periods between 584--747 days. Using a similar method as before, we see that the detection at 669 days (for the 36-day rebinned $F_{10.7}$ data) is unlikely to be associated with the $\frac{D}{5}$ overtone which we would expect around 792 days, as in this case the duration of the dataset is 3960 days. The detection at 747 days (from the 72-day rebinned $F_{10.7}$ data) is in the range of what we would expect for the $\frac{D}{5}$ overtone at 763 days, as the duration here is 3816 days. Although the $\frac{D}{5}$ overtone is not commonly observed, it is unlikely that the detection at 747 days is of solar origin and is more likely an artefact of the analysis method. We see another detection from the 72-day rebinned \ion{Mg}{ii} index data at 648 days. This detection is far from the expected range of the $\frac{D}{5}$ overtone at 864 days, as the duration is 4320 days for this dataset, and therefore we suggest this detection is not the result of overtones. The last detection in the QBO range comes from the 36-day rebinned \ion{Mg}{ii} index data with an average period of 584 days. The duration of this dataset is 4284 days, putting the $\frac{D}{5}$ (the overtone with the lowest periodicity that we usually observe) at 857 days. As the periodicity we detected is lower than this value, it's unlikely that this detection is an overtone and an artefact of the analysis method and more likely is of real solar origin. The final detection for Cycle~23 is at 1332 days from the 36-day rebinned $F_{10.7}$ data and closely matches up with the expected location of the $\frac{D}{3}$ overtone at 1320 days. 

Finally, we assess the results from Cycle~24. There is a detection of an IMF with a period of 448 days, originating from \ion{Mg}{ii} index data. This result is within the expected range for the annual oscillation but is higher in periodicity than those attributed to the annual oscillation in Cycles~22 and 23. This reflects the findings discussed in Section~\ref{sec:analysis} where for both GONG and MDI/HMI data we see an increase in detections of IMFs with periods shifted towards approximately 450 days in Cycle~24 compared to Cycle~23 (see the distribution of the histograms around 450 days in Cycles~23--24 in both Figures~\ref{fig:GONG},~\ref{fig:MDIHMI}). This shift in Cycle~24 may be due to some mixing between the lower-amplitude, higher periodicity QBO regime and the higher-amplitude, lower periodicity annual oscillation. All other oscillations in the results from Cycle~24 lie close to the expected location of some overtone. The detection at 961 days for the 36-day rebinned $F_{10.7}$ data is close to the location of the $\frac{D}{4}$ overtone at 945 days (where the duration of the dataset is 3780 days). The detections for the \ion{Mg}{ii} index datasets at 1246 and 1081 days for the 36-day and 72-day rebinned data respectively both lie close to the $\frac{D}{3}$ and $\frac{D}{4}$ overtones at 1089 and 1062 days where the durations of their datasets are 4356 and 4248 days respectively. Therefore we see that there is no strong evidence of QBO behaviour in Cycle~24 as observed by solar activity proxies. 

To summarise the results of both solar activity proxies excluding the IMFs that we can confidently attribute to overtones, we see oscillations of 464$\genfrac{}{}{0pt}{2}{+96}{-151}$ days in Cycle~21, 286$\genfrac{}{}{0pt}{2}{+167}{-149}$, 351$\genfrac{}{}{0pt}{2}{+88}{-100}$, 893$\genfrac{}{}{0pt}{2}{+163}{-221}$ days in Cycle~22, 308$\genfrac{}{}{0pt}{2}{+66}{-59}$, 584$\genfrac{}{}{0pt}{2}{+140}{-166}$, 648$\genfrac{}{}{0pt}{2}{+138}{-116}$ and 669$\genfrac{}{}{0pt}{2}{+173}{-179}$ in Cycle~23 and 448$\genfrac{}{}{0pt}{2}{+113}{-71}$ in Cycle~24.

For all solar activity proxy datasets, Fourier analysis did not detect any oscillations above a 95$\%$ confidence level other than the duration of the input signal. The annual oscillation was not seen above this level in any of the datasets which may be in part due to the fact that the majority of the power was distributed into longer period oscillations. Through detrending the signal first and then analysing it with FFT, it may be possible to uncover the annual oscillation. This may also reveal further information about the QBO although as the QBO exhibits non-periodic behaviour, power is likely to be distributed over a number of bins in frequency space, reducing its overall significance, and so the annual oscillation may still not be visible. This could be further investigated by rebinning the Fourier spectra \citep[see][for further details]{2017Pugh}. The Fourier spectra of $F_{10.7}$ index data over Cycle~22 can be seen in the lower left panel of Figure~\ref{fig:radio22_36}. 

In \citet[][]{2015Kolotkov}, a similar analysis was carried out on smoothed $F_{10.7}$ index data which was rebinned to both 108 and 10-day cadences and assessed with the Hilbert-Huang Transform (HHT). These analyses obtained a number of IMFs with periods that are consistent with our results. For example, both the 108-day and 10-day $F_{10.7}$ index revealed IMFs with periods of 1180$\genfrac{}{}{0pt}{2}{+337}{-214}$ and 1110$\genfrac{}{}{0pt}{2}{+710}{-340}$ days respectively which may be compared to the IMFs with weighted-average periods of 1081$\genfrac{}{}{0pt}{2}{+150}{-166}$ and 1246$\genfrac{}{}{0pt}{2}{+233}{-309}$ days detected in Cycle~24 in this study.

Similarly, IMFs with periods of 885$\genfrac{}{}{0pt}{2}{+117}{-200}$ and 833$\genfrac{}{}{0pt}{2}{+417}{-63}$ days found in the previous study can be compared to the detections with periods of 839$\genfrac{}{}{0pt}{2}{+86}{-110}$ and 893$\genfrac{}{}{0pt}{2}{+163}{-221}$ days in Cycle~22. \citet{2015Kolotkov} also observed IMFs firmly in the QBO range with periods of 708$\genfrac{}{}{0pt}{2}{+215}{-163}$ and 690$\genfrac{}{}{0pt}{2}{+80}{-200}$ days which may be compared to our detections in Cycle~23 of 747$\genfrac{}{}{0pt}{2}{+205}{-167}$ and 648$\genfrac{}{}{0pt}{2}{+138}{-116}$,669$\genfrac{}{}{0pt}{2}{+173}{-179}$ days respectively. These authors also detected a number of lower periodicity IMFs with periods of less than 200 days. However, the methodology used in that paper and our study have several differences which may be responsible for these additional IMFs. Additionally, the data used in \citet{2015Kolotkov} spanned over a continuous duration of two-and-a-half solar cycles whereas we use independent cycles which are trimmed to centre around periods of high solar activity. Finally, no significance testing was performed in the earlier study which naturally leads to a greater number of reported IMFs.

\section{Conclusion}
\label{sec:conclusion}

We have used both EMD and Fourier analyses to study quasi-periodicities in the time variation of frequency shifts of p-modes observed with GONG and MDI/HMI. The results in this paper were confined to data that had sufficiently low error, restricting the modes from which we found significant IMFs to roughly 2600--3400\,$\mu$Hz and 0.6--0.9\,R$_{\sun}$. We also examined the $F_{10.7}$ and \ion{Mg}{ii} index over four solar cycles to search for the QBO.  Our paper concludes the following. \\

1.\textbf{ We find evidence of the QBO in the combined durations of Cycles~23 and 24 in both GONG and MDI/HMI data} (see Figure~\ref{fig:combined2324}). We observe periodicites in three main groups for GONG data at 500--800, 900--1200, 1200--1500 days and in two groups for MDI/HMI data at 600--1000 and 1200--1500 days which we consider to be evidence of the QBO. \\

2.\textbf{ We also observe evidence of the QBO in Cycles~ 23 and 24 when analysed separately. These observations are accompanied by overtones (specifically with periods in the range of $\frac{D}{3}$, $\frac{D}{4}$ and possibly $\frac{D}{5}$). The ranges where potential QBO candidates were observed overlap with periods where overtones are expected making a definitive classification difficult.}
The observed periodicities not solely attributed to artefacts or overtones were approximately between 500--800 days for Cycle~23 across both datasets. MDI/HMI data revealed periodicities over a similar range of 550-800 days in Cycle~24 which we attribute to the QBO. Conversely GONG data did not produce strong evidence of the QBO in Cycle~24, where the majority of IMFs overlapped with a region associated with an overtone making it difficult to ascertain whether these IMFs are an artefact of the analysis procedure or true QBO candidates. 
The EMD technique uncovered consistent measurements of $\frac{D}{3}$ and $\frac{D}{4}$ overtones at approximately 1300 and 1000 days for GONG data over both cycles. Similarly, MDI/HMI datasets yielded detections of $\frac{D}{3}$, $\frac{D}{4}$ and $\frac{D}{5}$ overtones with periods of approximately 1200, 900, and 700 days across both cycles. \\

3.\textbf{ We see a weaker presence of the QBO in Cycle~24 compared to Cycle~23 which we do not attribute to data sparsity or analysis methods.} This is because we see fewer IMFs in the approximate QBO range from 500--800 days in Cycle~24 than Cycle~23 (shown in the lower panels of Figures~\ref{fig:GONG} and \ref{fig:MDIHMI}), despite the number of common modes increasing for GONG data (by 57$\%$) and remaining roughly constant for MDI/HMI data (an increase of 3$\%$). 

We further note that the increase in common modes between Cycles 23 and 24 for GONG data has not changed the overall number of detections of the annual oscillation, although Cycle~24 shows significantly more scatter in the 250--450 range compared to the tight banding seen in Cycle~23. This implies that an oscillation of physical origin is still visible in Cycle~24 data which contrasts with the sparsity of IMFs corresponding to the QBO range. The annual oscillation is not observed by MDI/HMI, however, we see the same trend in data from both instruments - a lower population of IMFs with periods between 500--800 days in Cycle~24 when compared to their Cycle~23 counterparts. 

The relative significance of the overtone oscillations at $\frac{D}{3}$,$\frac{D}{4}$ (which could be mistakenly attributed to high periodicity QBO candidates) were largely unaffected between the cycles, whilst the apparent presence of the QBO candidates reduced. This may lead a reader to the incorrect interpretation that the periodicity of QBO increased between Cycles~23 and 24.

Since the weaker signature of the QBO in Cycle~24 is seen in two different datasets and does not depend on the number of common modes, we believe that this can not be attributed to the analysis procedure.  The reduced presence of the QBO in Cycle~24 further suggests that the QBO is a cycle-dependent process and its generation mechanism is in some way linked with that of the Schwabe cycle.\\

4. \textbf{There is some evidence of the QBO in solar activity proxies, specifically during Cycles~22 and 23. This may suggest some form of correlation between the observational properties of the Schwabe cycle and the QBO.}
We observe QBO candidates with periods of around 450 days in Cycle~21 and 24, approximately 900 days in Cycle~22 and around 650 days in Cycle~23. Significant IMFs with periods that corresponded to overtones and Earth's annual orbit were also uncovered. The detection at 450 days in Cycle~24 is interesting as that was also seen in the MDI data, and potentially in the GONG data, although here it is hard to distinguish from the annual periodicity. 
We point out that the periodicities detected in the QBO range occurred during the more active solar Cycles; Cycles~22 and ~23. This may suggest a that the amplitude of the QBO in both frequency shift data and solar activity proxy data scales with the activity of the solar cycle.  \\

5. \textbf{The presence of the QBO was not affected by the depth to which the p-mode travelled, nor the average frequency of the p-mode. The analysis further suggests that the magnetic field responsible for producing the QBO in frequency shifts of p-modes is anchored above approximately 0.95~R$_{\sun}$.}
The QBO candidates seen in Figures~\ref{fig:combined2324}, \ref{fig:GONG} and \ref{fig:MDIHMI} are present across almost all depths where data is available. There is no dependence with lower turning points of any of the modes with the appearance of the QBO, other than in Cycle~24 for MDI/HMI data where the 700--800 and 800--900 days bands are more densely populated at radii greater than 0.75\,R$_{\sun}$. As the intermediate degree modes penetrate to all depths, this suggests that the magnetic field generating the QBO must be anchored above the highest lower turning point of any mode which shows QBO behaviour. The upper turning points of these modes could be analysed in order to better determine the location of the magnetic field.\\

6. \textbf{There is difficulty differentiating overtones and the QBO due to spread in periodicity and large errors.}
The errors associated with the periods of statistically significant IMFs are large (see Figures~\ref{fig:GONG}, \ref{fig:MDIHMI}), with upper and lower bounds overlapping across several bands. Neglecting errors, we find that 20$\%$ of IMF periods fall within the expected overtone ranges (visualised by pink shaded bands) for GONG data in Cycle~23 (increasing to 30$\%$ for Cycle~24). For MDI/HMI data the percentages of IMFs that have a period within the ranges of overtones are 20$\%$ and 58$\%$ for Cycles~23 and 24 respectively. Including errors, most of the QBO periods overlap with at least one overtone band, making it even more challenging to assess if a statistically significant period is the result of an overtone. 
However we do expect the errors on these IMFs to be large due to the nature of quasi-periodic signals. Recall that the errors on the average periods of these IMFs are obtained by the full width half maximum of the IMF's global wavelet spectrum. Therefore a signal with significant period drift will naturally produce large errors. The contrast of this can be seen by examination of the IMFs attributed to the annual oscillation (which has a stationary period) which show considerably smaller errors than the QBO candidates. We also expect the presence of overtones which are more frequently observed in shorter duration data. This can be seen in comparing the presence of overtones in single cycle data (Figures~\ref{fig:GONG}, \ref{fig:MDIHMI}) to the longer duration combined cycled data (Figure~\ref{fig:combined2324}).  
Although the large errors and presence of overtones are to be expected, the combination of the two leads to difficulty in interpreting the origin of IMFs. At present there is no technique by which we can reliably identify which regime an individual IMF belongs. This can be mitigated by addressing the behaviour of clusters of IMFs instead of individual IMFs but is beyond the scope of the present paper. \\

7. \textbf{The Fast Fourier Transform is poorly suited to assessing signals with significant period drift.}

We assessed helioseismic data using the Fast Fourier Transform and rarely found oscillations with periods in the QBO range. Similarly, we found no evidence of the QBO when assessing solar activity proxies with the Fast Fourier Transform. We believe this is mainly due to spectral leakage between adjacent period bins in the power spectra of the signals, due to the period drift present in QBO signals. In order to properly analyse quasi-oscillatory signals, we must ensure we make use of techniques capable of handling period-drifting signals for quasi-oscillatory processes such as EMD or wavelet analysis. \\

Helioseismic data spanning a larger time range would improve this study as it would not only allow us to look at the longer-term evolution and behaviour of the oscillations in the QBO range, but would also reduce the presence of overtones, which have an inverse dependence on the duration of the input signal. Therefore we raise the importance of long-term synoptic helioseismic datasets, such as GONG and HMI, which is needed if we hope to study the temporal evolution of phenomena confined to or generated in the solar interior.  It would be of interest to the community to revisit this study following Cycle~25.

\section*{Acknowledgements}
We thank the anonymous referee for helpful comments which have improved the contents of the paper. TM would like to thank the National Solar Observatory (NSO) for hospitality, where a part of this work was carried out. This research data leading to the results obtained has been supported by SOLARNET project that has received funding from the European Union’s Horizon 2020 research and innovation programme under grant agreement No. 824135. D.Y.K. and A-MB acknowledge support from the Science and Technology Facilities Council (STFC) consolidated grant ST/T000252/1.
RK and A-MB recognise the support of STFC consolidated grant ST/P000320/1. This work utilises GONG data obtained by the NSO Integrated Synoptic Program, managed by the National Solar Observatory, which is operated by the Association of Universities for Research in Astronomy (AURA), Inc. under a cooperative agreement with the National Science Foundation and with contribution from the National Oceanic and Atmospheric Administration. The GONG network of instruments is hosted by the Big Bear Solar Observatory, High Altitude Observatory, Learmonth Solar Observatory, Udaipur Solar Observatory, Instituto de Astrof\'{\i}sica de Canarias, and Cerro Tololo Interamerican Observatory. This work utilises data from the SoHO/MDI and SDO/HMI. SoHO is a mission of international cooperation between ESA and NASA. SDO data courtesy of SDO (NASA) and the HMI and AIA consortium. The MDI and HMI data are obtained from the Joint Science Operations Center at Stanford University. We also thank the University of Bremen, Bremen, Germany, and National Research Council Canada (in partnership with the Natural Resources Canada). TM would also like to thank C. E. Pugh for helpful discussions and S. Afinogenov for the co-development of the EMD code on which this work is based. This work also made use of the following software libraries: NumPy \citep{numpy}, MatPlotLib \citep{matplotlib}, SolarSoft \citep{1998solarsoft}, and SciPy \citep{scipy}.

\section*{Data availability}
The data used here are all publicly available. The helioseismic frequency datasets and activity indices can be downloaded from: GONG (\url{https://gong2.nso.edu/}),  MDI (\url{http://jsoc.stanford.edu/MDI/MDI\_Global.html}), HMI  (\url{http://jsoc.stanford.edu/HMI/Global\_products.html}),  DRAO Adjusted $F_{10.7}$ index (\url{https://www.spaceweather.gc.ca/solarflux/sx-5-en.php}) and \ion{Mg}{II} index (\url{http://www.iup.uni-bremen.de/UVSAT/Datasets/MgII}). 




\bibliographystyle{mnras}
\bibliography{qbobib} 




\bsp	
\label{lastpage}
\end{document}